\documentclass[aps,prb,reprint,groupedaddress,floatfix]{revtex4-1}
\usepackage{hyperref}
\usepackage{subeqnarray}
\usepackage{comment}
\usepackage{amsmath,amssymb,bbm}
\usepackage{graphics,graphicx}
\usepackage{mathrsfs}
\usepackage{dsfont}
\usepackage{cleveref}
\usepackage[dvipsnames]{xcolor}

\newcommand{\N}  {M}
\newcommand{\Ne} {N}
\newcommand{\Nb} {\bar{N}}
\newcommand{\cn}   {c_n}
\newcommand{\cnb} {\bar{c}_n}
\newcommand{\sign}  {\trm{sign}}

\newcommand{\Dmgs} {\mathcal{D}}
\newcommand{\Dmgsmu} {\mathcal{D}_{\mu}}

\newcommand{\Dmhp} {D^{+}}

\newcommand{\Hmgs} {\mathcal{H}}

\newcommand{\Egs}  {\mathcal{E}}
\newcommand{\Dminf}{D_{\infty}}
\newcommand{\tHmax} {\widetilde{\mathcal{H}}_{\trm{max}}}
\newcommand{\tHmin}  {\widetilde{\mathcal{H}}_{\trm{min}}}

\renewcommand{\Im} {I}
\newcommand{\minimize}[1]{\underset{#1}{\operatorname{minimize}}\:}
\newcommand{\Dm}{D}
\newcommand{\Db}{\bar{D}}

\newcommand{\pb}{\bar{p}}

\newcommand{\I}{I}

\newcommand{\thetab}  {\bar{\theta}}
\newcommand{\tmu }{\widetilde{\mu}}

\newcommand{\trace} [1]{\trm{Tr}\{#1\}}
\newcommand{\xph}{x_{\bar{p}}}
\newcommand{\xpp}{x_p}
\newcommand{\xpm}{x_m}

\newcommand{\Trace} [1]{\trm{Tr}\{#1\}}
\newcommand{\Fnorm} [1]{\|#1\|_\mathcal{F}}

\newcommand{\norm}  [1]{\|#1\|}
\newcommand{\diag}  [1]{\trm{diag}\{#1\}}

\newcommand{\homo}{\epsilon_{\Ne}   }
\newcommand{\lumo}{\epsilon_{\Ne +1}}
\newcommand{\gap} {\Delta\epsilon_\trm{gap}}
\newcommand{\tbeta}{\beta}
\newcommand{\bbeta}{\bar{\beta}}

\newcommand{\FuncMcW}[2]{\Omega_{\text{McW}}\{#1\}_{#2}}

\newcommand{\LagMcW}[2]{\mathcal{L}_{\text{McW}}\{#1\}_{#2}}
\newcommand{\GradLag}[2]{\nabla\mathcal{L}_{\text{McW}}\{#1\}_{#2}}

\newcommand{\funcMcW}{\Omega_{\text{McW}}}
\newcommand{\gradfunc}{\nabla\funcMcW}
\newcommand{\lagMcW}{\mathcal{L}_{\text{McW}}}
\newcommand{\gradlag}{\nabla\lagMcW}

\newcommand*{\ie}{ie.}

\newcommand*{\trm}{\textrm}
\newcommand*{\eq}  [1]{Eq.~(\ref{#1})}

\newcommand*{\fig} [1]{Fig.~\ref{#1}}
\newcommand*{\reff}[1]{(\ref{#1})}
\newcommand*{\citen}[1]{Ref.~[\onlinecite{#1}]}

\begin{document}

\title{Generalized canonical purification for density matrix minimization}

\author{Lionel A. Truflandier}
\email{lionel.truflandier@u-bordeaux.fr}

\author{Rivo M. Dianzinga}
\affiliation{Institut des Sciences Mol\'{e}culaires, Universit\'{e} Bordeaux,
CNRS UMR 5255, 351 cours de la Lib\'{e}ration, 33405 Talence cedex, France}

\author{David R. Bowler}
\affiliation{London Centre for Nanotechnology, UCL, 17-19 Gordon St, London WC1H 0AH, UK}
\affiliation{Department of Physics \& Astronomy, UCL, Gower St,
  London, WC1E 6BT, UK}
\affiliation{International Centre for Materials Nanoarchitechtonics
  (MANA), National Institute for Materials Science (NIMS), 1-1 Namiki,
  Tsukuba, Ibaraki 305-0044, JAPAN}

\date{\today}
\begin{abstract}
A Lagrangian formulation for the constrained search for the $N$-representable one-particle 
density matrix based on the McWeeny idempotency error minimization is proposed, which converges 
systematically to the ground state.  A closed form of the canonical purification is derived for which 
no \textit{a posteriori} adjustement on the trace of the density matrix is needed.  The relationship
with comparable methods are discussed, showing their possible generalization through the 
\textit{hole-particle} duality. The appealing simplicity of this \textit{self-consistent} recursion 
relation along with its low computational complexity could prove useful as an alternative to 
diagonalization in solving dense and sparse matrix eigenvalue problems.
\end{abstract}
%
%
\maketitle
As suggested 60 years ago\cite{McWeeny_density_1956a,*McWeeny_density_1956b,*McWeeny_density_1957c},
the idempotency property of the density matrix (DM) along with a minimization algorithm
would be sufficient to solve for the electronic structure without relying on the time consuming 
step of calculating the eigenstates of the Hamiltonian matrix. The celebrated McWeeny purification 
formula\cite{McWeeny_RevModPhys_1960} has inspired major advances in electronic structure 
theory based on (conjugate-gradient) density matrix minimization\cite{LNV_PRB_1993,Daniels_JCP_1997,Millam_JCP_1997,Daniels_JCP_1999,Bowler_LNV_1999,Challacombe_JCP_1999}
(DMM), or density matrix polynomial expansion\cite{Goedecker_PRL_1994,Goedecker_PRB_1995} 
(DMPE), where the 
density matrix is evaluated by the recursive application of projection 
polynomials (commonly referred as \textit{purification}). DMPE resolution 
includes  the  Chebyshev polynomial recursion\cite{Goedecker_PRL_1994,Goedecker_PRB_1995,Baer_PRL_1997,Baer_JCP_1997,Bates_JCP_1998,Liang_JCP_2003,Niklasson_PRB_2003}, the Newton-Schultz $\sign$ matrix iteration\cite{Nemeth_JCP_2000,Beylkin_SIGN_1999,Kenney_sign_1991},
the trace-correcting\cite{Niklasson_TCP_2002}, trace-resetting\cite{Niklasson_TRS_2003} 
purification (TCP and TRS, respectively), and the Palser and Manolopoulos canonical purification (PMCP)\cite{PM_PRB_1998}.
They constitute, with sparse matrix algebra, the principal ingredient for efficient
linear-scaling tight-binding (TB) and self-consistent field (SCF) theories\cite{Bowler_ordern_2012,Goedecker_RevModPhys_1999}.
Unfortunately, since all these methods were originally derived within the grand canonical ensemble\cite{DFT_Bible_1994}, for a given a total number of states ($M$), 
none of them are expected to yield the correct number of occupied states ($N$) unless the chemical 
potential ($\mu$) is known exactly. As a result, solutions rely on heuristic 
considerations, where the value of $\mu$\cite{Baer_JCP_1997}, or 
the polynomial expansion\cite{Niklasson_TCP_2002} are adapted 
\textit{a posteriori} to reach the correct value for $N$, which add
irremediably to the computational complexity.
Despite the remarkable performances of the DMPE approaches for solving for sparse\cite{Rudberg_JPCM_2011,Daniels_JCP_1999} and dense\cite{Chow_JCP_2015,Cawkwell_JCTC_2012,Cawkwell_JCTC_2014} DM, 
they remain unsatisfactory since they constitute a formalism which does not account
explicitly for the canonical requirement of constant-$N$. 

\begin{figure}[!hptb]
  \centering
  \includegraphics[scale=0.37]{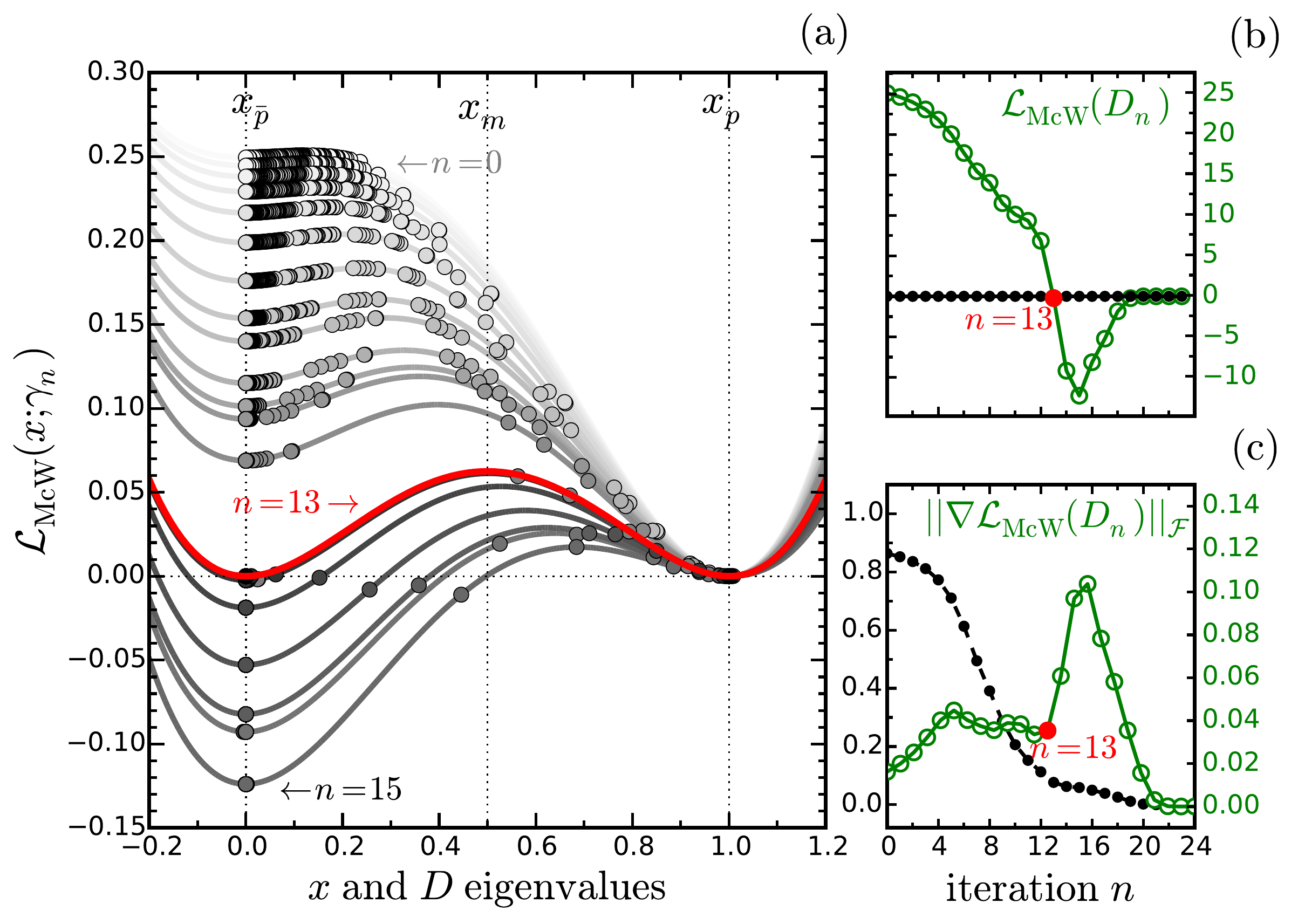}  
  \caption{
    (a) Convergence of the McWeeny Lagrangian and density matrix eigenvalues
    during the course of the minimization using a test Hamiltonian and an occupation factor 
    $\theta=0.10$. A grey scale is used to guide the eye during the processus of purification. 
    Each curve is a plot of the function 
    $\lagMcW(x;\gamma_n)$ computed at each iteration $n$. 
    The red line correspond to $\lagMcW(x;0) = \funcMcW$. 
	(b) Convergence of $\lagMcW$ (green circles), 
    and the trace conservation $\Trace{\Dm_n} - \Ne$ (black dots). 
    (c) Convergence of $\Fnorm{\gradlag}$ (green circles) 
     and $\Fnorm{\Dm_n}-\Ne$ (black dots). 
    } 
  \label{figure1}  
\end{figure}

In this letter, we derive a rigorous and variational constrained search 
for the one-particle density matrix which does not rely on \textit{ad hoc} adjustments and respects 
the $N$-representability constraint throughout the minimization (or purification) process.  
We shall start from the McWeeny unconstrained minimization of the error in
the idempotency of the density matrix\cite{McWeeny_density_1956b}, given by
\begin{subequations}
\label{eq:mcweeny1}
\begin{align}
  \minimize{\Dm\rightarrow\Dmgsmu}
   &\ \FuncMcW{D;(\Hmgs,\mu)}{}   \label{eq:mcw1a}\\
   \trm{with:}\ \ \funcMcW &= \Trace{(\Dm^2 - \Dm)^2}\label{eq:mcw1b}
\end{align}
\end{subequations}
where, for a given fixed Hamiltonian\footnote{We will restrict our discussion to effective 
one-electron Hamiltonian operators expressed in a finite orthonormal Hilbert space. 
It does not pose severe challenge to generalize the demonstration to
a non-orthogonal basis set.} $\Hmgs$ and chemical potential $\mu$, 
the density matrix $\Dmgsmu$ is the ground-state for that Hamiltonian
and chemical potential. The initial guess for $D$ is generally  constructed 
as a function $\Hmgs$, suitably scaled:
\begin{equation}
\label{eq:dmguess} 
	   \Dm_0 =  \beta_1\Im + \beta_2(\mu\Im-\Hmgs) 
\end{equation}
where $\beta_1$ and $\beta_2$ stand for preconditioning constants such that the 
eigenvalues of $\Dm_0$ lie within a predefined range.
The double-well shape of the McWeeny function with 3 stationnary points: 2 minima at 
$\xpp=1$ and $\xph=0$, and 1 local maximum at $\xpm=\frac{1}{2}$ 
(see~\fig{figure1}a, red curve), are important features in developing robust  DMM algorithms. 
Finding the minimum of $\funcMcW$ would be easily performed by stepwise gradient 
descent\cite{McWeeny_density_1956a}, where the density matrix is updated at each 
iteration $n$,
\begin{subequations}
\label{eq:mcweeny2}
\begin{align}
   \Dm_{n+1}&=\Dm_{n} - \sigma_{n}\gradfunc\label{eq:mcw2a}\\
   \trm{with:}\ \ \gradfunc &= 2 \left( 2 \Dm_n^3 - 3\Dm_n^2 + \Dm_n\right)\label{eq:mcw2b}
\end{align}
\end{subequations}
and $\sigma_{n} \geq 0$ represents the step length in the negative direction of the gradient.
Considering an optimal fixed step length descent ($\sigma=1/2$), on inserting \eq{eq:mcw2b} 
into \eq{eq:mcw2a}, the McWeeny purification formula appears
\begin{equation}
\label{eq:mcweeny3}
  \Dm_{n+1} = 3\Dm_{n}^2 - 2\Dm_{n}^3
\end{equation}
where the right-hand-side of the equation above can be view as an auxiliary DM.
For a well-conditioned $\Dm_0$, \ie\ $\lambda(\Dm_0)\in[-\frac{1}{2},\frac{3}{2}]$, 
repeated application of the recursion identity [\eq{eq:mcweeny3}] naturally drives 
the eigenvalues of $\Dm_{n+1}$ towards 0 or 1. For basic TB Hamiltonians where 
the occupation factor ($\theta=N/M$) is close to $1/2$ and $\mu$ can be 
determined by symmetry\cite{PM_PRB_1998}, or when the input DM is already 
strongly idempotent, the minimization principle \reff{eq:mcw1a} is able, on its own, 
to deliver the correct $N$-representable $\Dmgs$.  Beyond these very specific cases, 
we have to enforce the objective function \reff{eq:mcw1b} 
to keep $\Ne$ constant during the minimization.  From \eq{eq:mcweeny3},
a sufficient condition would be to impose the trace of the auxiliary DM to
give the correct number of occupied states.
This leads us to solve a constrained optimization problem which can be formulated in terms 
of the McWeeny Lagrangian, $\lagMcW$, using the following minimization principle:
\begin{subequations}
\label{eq:mcweeny4}
\begin{align}
	\underset{
	\begin{subarray}{c}	
	\{\Dm\rightarrow\Dmgs|\Trace{\Dm}=\Ne\}\\ 
	\gamma
	\end{subarray}}{\operatorname{minimize}}\:& 	
	\ \LagMcW{D,\gamma;(\Hmgs),\Ne}{} \label{eq:mcw4a} \\
   \trm{with:}& \nonumber\\
   \lagMcW   =  \funcMcW &- \gamma\left( \Trace{3\Dm^2 - 2\Dm^3} - \Ne \right)     	
   \label{eq:mcw4b} 
\end{align}
\end{subequations}
where $\gamma$ is the constant-$\Ne$ Lagrange multiplier. 
The McWeeny Lagrangian can be minimized using any gradient-based 
methods, with:
\begin{subequations}
\label{eq:mcweenygrad}
\begin{align}
   \gradlag   &=  \gradfunc  - 6\gamma\left(\Dm - \Dm^2 \right)    \label{eq:mcwgrada}\\
   \partial_\gamma\lagMcW &=  \Trace{3\Dm^2 - 2\Dm^3} - \Ne   \label{eq:mcwgradb}
\end{align}
\end{subequations}
Taking the trace \eq{eq:mcwgrada}, and inserting the constraint of \reff{eq:mcwgradb}, 
we obtain the expression for $\gamma$:
\begin{subequations}
\label{eq:mcweenygamma}
\begin{align}
	 \gamma         & = \frac{1}{3} - \frac{2}{3}c  -  \frac{1}{6}d \label{eq:mcwgama}\\
	  \trm{with:}\ c&= \frac{\Trace{\Dm^2 - \Dm^3}}{\Trace{\Dm - \Dm^2}}\label{eq:mcwgamb}\\
	                     d&= \frac{\Trace{ \gradlag}}{\Trace{\Dm - \Dm^2}} \label{eq:mcwgamc}
\end{align}
\end{subequations}
Then, by substituting \eq{eq:mcwgrada} in \eq{eq:mcwgamc}, we can easily 
show that $d=0$, that is $\Trace{\gradlag}=0$, $\forall\Dm$. As a result, 
given $\Dm_0$ such that $\Trace{\Dm_0}=\Ne$, and the fixed-step 
gradient descent minimization, we obtain a recursion formula:
\begin{equation}
	  \Dm_{n+1} = \Dm_{n} - \frac{1}{2}\GradLag{\Dm_n;\gamma_n}{} \label{eq:hpcp1a}
\end{equation}
which guarantees $\Trace{\Dm_{n+1}}=\Ne$,  $\forall n$.
The parameter $c$ [\eq{eq:mcwgamb}] is recognized as the unstable fixed 
point introduced in \citen{PM_PRB_1998}, where $c\in[0,1]$. As a result, 
the interval $[-\frac{1}{3},\frac{1}{3}]$ constitutes the stable variational 
domain of $\gamma$. 

The variation of the McWeeny Lagrangian function and the density matrix eigenvalues 
during the course of the minimization are presented in \fig{figure1}a 
for a test Hamiltonian with $N=10$, $M=100$, and 
a suitably conditioned initial guess (\textit{vide infra}). The 
corresponding convergence profiles of $\lagMcW$ and $\norm{\gradlag}$ 
(green circles) are reported on Figs~\ref{figure1}b and \ref{figure1}c, respectively, 
along with the trace conservation $\Trace{\Dm_n}-\Ne$, and the  density
matrix norm convergence $\norm{\Dm_n}-\Ne$ (black dots).
We may notice first that for $\gamma=0$ (or $c=\xpm=\frac{1}{2}$), 
$\lagMcW$ simplifies to $\funcMcW$. For intermediate states, 
$\gamma\in]-\frac{1}{3},0[\cup]0,\frac{1}{3}[$, the symmetry of $\funcMcW$ 
is lost and the shape of $\lagMcW(x,\gamma_n)$ drives the eigenvalues in the 
\textit{left} or in the \textit{right} well. We may call them
the \textit{hole} and \textit{particle} well, respectively.
From the grey scale in \fig{figure1}a, we observe how $\gamma_n$ 
influences $\lagMcW$ (along the $y$-axis) at $\xph$, and the 
abscissa of the second stationnary point $\xpm$  which is free to move in 
$[\xph,\xpp]$. This yields to transform the \textit{hole} well from a local $(n=0)$
to a global $(n=15)$ minima (or conversely the \textit{particule} well from a global to 
a local minima). At the boundary values $\gamma=\{-\frac{1}{3},\frac{1}{3}\}$, $\xph$ 
and $\xpm$ merged to a saddle point in such a way that only one global minima 
left at $\xpp$. Notice that, for situations  where $\gamma\notin[-\frac{1}{3},\frac{1}{3}]$, 
the saddle point transforms to a maximum and runaway solutions may appear. 
Nevertheless, as long as $\Dm_0$ is well conditioned, such kind of critical problem 
should not be encountered.

Figs.~\ref{figure1}b and \ref{figure1}c highlight the minimization
mechanism: (i) From iterate $n=0$ to $12$; $\gamma\rightarrow 0^+:$
$\lagMcW$ follows the search direction and decreases monotonically.
(ii) At iterate $n=13$; $\gamma\simeq 0:$ $\lagMcW$ is close to the target value
but the gradient residual is nonzero. (iii) From $n=14$ to 15; $\gamma < 0:$ 
the search direction is inverted. (iv) At iterate $n=16$: all the eigenvalues are 
trapped in their respective wells. (iii) From iterate $n=17$ to $23$, $\gamma\rightarrow 0^-:$
we are in the McWeeny regime [\eq{eq:mcweeny3}] and $\lagMcW$ 
eventually reaches the global minimum.

\begin{figure*}[!ht]
  \centering
  \includegraphics[scale=0.46]{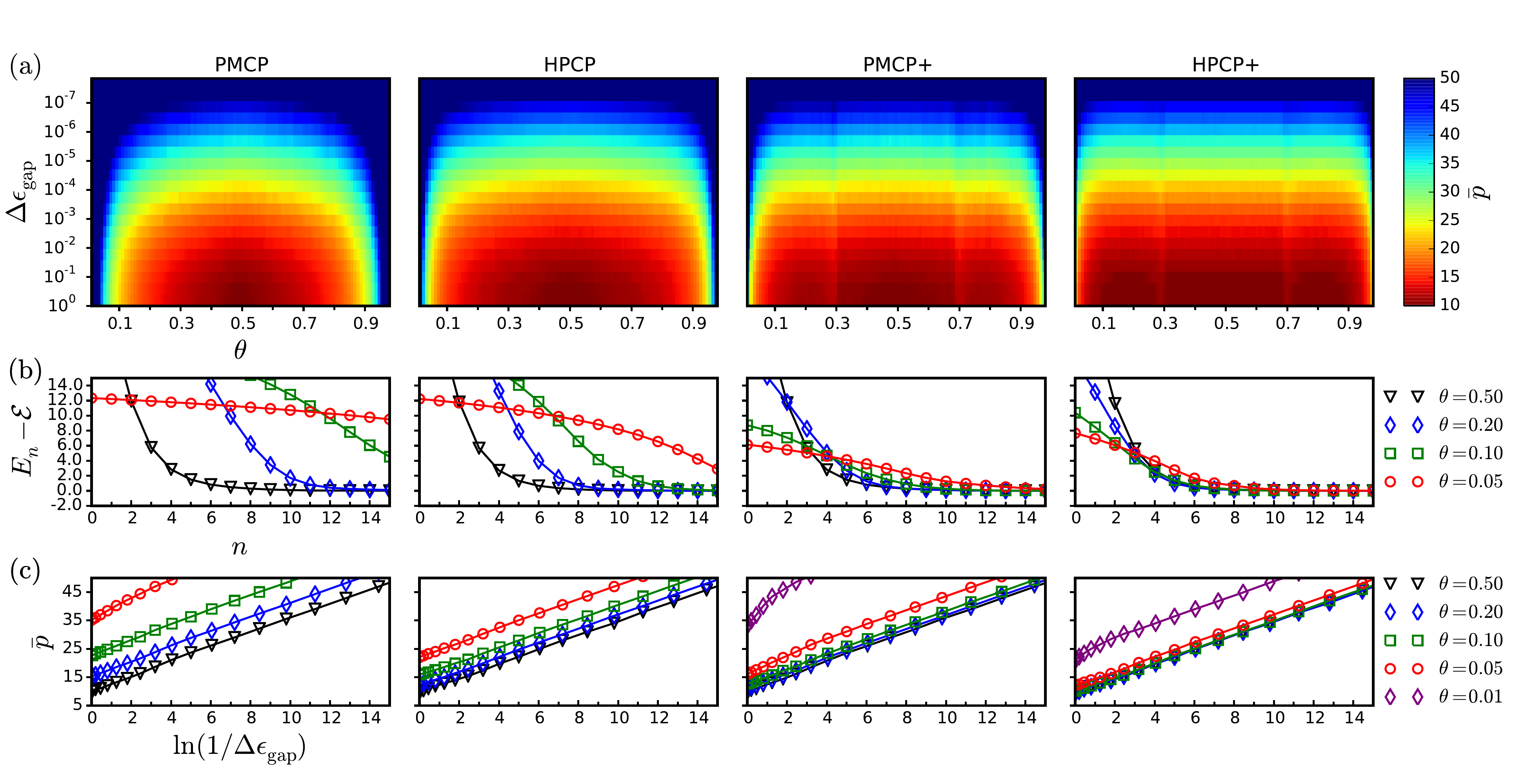}  
  \caption{(a) Color maps displaying the average number of purifications ($\pb$) as the function 
  of the filling factor ($\theta$) and energy gap ($\gap$). 
  Results obtained from the PMCP and HPCP methods using the initial guess 
  of Eqs.~\reff{eq:dmguess}-\reff{eq:pmcpguess},
  and  Eqs. \reff{eq:dmguess}-\reff{eq:hpcpguess} (notated  PMCP+ and HPCP+).
  Each pixel on the maps correspond  to an average over $32$ test Hamiltonians. 
  (b) Energy convergence profiles with respect to the first 15 iterations for selected values 
  of $\theta$. (c) Average number of purifications as a function of $\ln(1/\gap)$.}
  \label{figure3}
\end{figure*}

Taking advantage of the closure relation,
\begin{equation}
\label{eq:closure}
	\Db + \Dm = \Im
\end{equation}
where $\Db$ stands for the \textit{hole} density matrix\cite{Mazziotti_PRE_2003}, a more 
appealing form for the McWeeny canonical purification [\eq{eq:hpcp1a}] can be derived  by 
reformulating Eqs.~\reff{eq:mcwgrada} and~\reff{eq:mcwgamb} in terms of $\Dm$ and $\Db$:
\begin{equation}
\label{eq:hpcp3}
  \Dm_{n+1} = \Dm_{n} + 2 \left( \Dm^2_{n}\Db_{n} 
  - \frac{\Trace{\Dm^2_n\Db_n}}{\Trace{\Dm_n\Db_n}}\Dm_{n}\Db_{n}\right)
\end{equation}
Notice that since at convergence $\Dm\Db=0$, $\Trace{\Dm\Db}$ must be chosen 
as the termination criterion in the recusion of \eq{eq:hpcp3} to avoid numerical instabilities 
when approaching the minima. The closed-form of this recurrence relation is remarkable:  
providing $\Hmgs$ used to build $\Dm_0$ [\eq{eq:dmguess}] and $\Ne$,
we have a self-consistent purification transformation which should converge 
to $\Dmgs$ without any support of heuristic  adjustements. 
Indeed, \eq{eq:hpcp3} can also be derived from the PMCP relations by working on both
$\Dm$ and $\Db$, and enforcing relation \reff{eq:closure} at each iteration (see Appendix).
Consequently, we can also demonstrate\cite{MRD_unpublished} that the \textit{hole-particle} 
canonical purification (HPCP) of \eq{eq:hpcp3} converges quadratically on $\Dmgs$ as
shown on \fig{figure3}c.

To assess the efficiency and limitations of the HPCP, we have investigated the dependence of the
number of purifications ($p$) on the occupation factor ($\theta$), and the energy gap 
$(\gap=\lumo-\homo)$, defined by the higher-occupied  ($\homo$) and lower-unoccupied 
($\lumo$) states. Similarly to the protocol of 
Niklasson\cite{Niklasson_TCP_2002,Niklasson_PRB_2003}, sequences of $\N\times\N$ 
dense Hamiltonian matrices ($\N=100$) with vanishing off-diagonal elements were generated,
having eigenvalues randomly distributed in the range $[-2.5,\homo]\cup[\lumo,2.5]$
for various $\gap\in[10^{-7},1.0]$. As a first test, results are compared to the
PMCP\cite{PM_PRB_1998}, along with the original initial guess [\eq{eq:dmguess}], 
where:
\begin{eqnarray}
\label{eq:pmcpguess}
	\beta_1 = \theta &,&\quad
    \beta_2 = \min\left\lbrace\tbeta,\bbeta\right\rbrace \label{eq:pmcpguessa}\\
    \trm{with:}\quad\quad\quad\quad\nonumber   & &\\
        \tbeta    =  \frac{\theta }{\tHmax - \mu},    & &\quad
        \bbeta   =  \frac{\thetab}{\mu -  \tHmin},    \quad 
        \mu\simeq\tmu=\frac{\trace{\Hmgs}}{\N}  \nonumber
\end{eqnarray}
and $\thetab = 1 - \theta = \Nb/\N$, $\Nb$ being the number of 
unoccupied states. The lower and upper bounds of the Hamiltonian eigenspectrum 
($\tHmin$ and $\tHmax$, respectively) were estimated from to the Ger{\v s}gorin's
 disc theorem\cite{Gerschgorin_disc_1931}. The preconditioning of $\Dm_0$ given in 
\eq{eq:pmcpguessa} guarantees that $\lambda(\Dm_0)\in[0,1]$, and gives 
rise to the following additional constraints: 
\begin{subequations}
\label{eq:pmcpguessconstraints}
\begin{align}
	\Trace{\Dm_0}     & =   \Ne \label{eq:pmcpcons1} \\
	\Trace{\Dm_0}     & >   \Trace{\Dm^2_0} > \Trace{\Dm^3_0} \label{eq:pmcpcons2}\\
	\Trace{\Dm^3_0} & > 2\Trace{\Dm^2_0} - \Trace{\Dm_0}      \label{eq:pmcpcons3}
\end{align}
\end{subequations}
which are also necessary and sufficient conditions for $c\in[0,1]$ at the first iteration. 
Convergence was achieved with respect to the idempotency property, 
such that $\Trace{\Dm_n\Db_n} \leq 10^{-6}$ for all the calculations.
Additional tests on the Frobenius norm\footnote{The Frobenius norm is defined by:
$||\Dm||_\mathcal{F} = (\sum_{i,j}|D_{ij }|^{2})^{1/2} = \sqrt{\Trace{\Dm^2}}.$ 
Notice that:  $\forall\Dm$, such that $\Dm^2=\Dm$, then 
$||\Dm||_\mathcal{F} = \sqrt{\Trace{\Dm}}$.} and the eigenvalues 
of the converged density matrix ($\Dminf$) were performed, using:
\begin{subeqnarray}
\slabel{eq:testa}
  \Fnorm{\Dminf}  &-& \sqrt{\Trace{\Dminf}} < 10^{-6}\\
\slabel{eq:testb}  
  \Fnorm{\Dminf}  &-& \Ne  < 10^{-6} \\
\slabel{eq:testc}  
     \Fnorm{\diag{\Dminf}  &-& \diag{\Im_{\Ne},0_{\Nb}}}  < 10^{-6} 
\end{subeqnarray}
which ensures that, at convergence, the representation of $\Dminf$ is  orthogonal, 
and $\Dminf$ corresponds to the exact $N$-representable ground-state density
matrix $\Dmgs$.

The variation of the average number of purifications ($\pb$) with respect to $\theta$ 
and $\gap$ are displayed on \fig{figure3}a using a color map for $\pb\in[10,50]$. 
For a given energy gap, the HPCP shows a net improvement over the PMCP approach 
regarding moderate low and high occupation factors. Nevetheless, as previously noted by 
Niklasson and Mazziotti\cite{Niklasson_TCP_2002,Mazziotti_PRE_2003}, the extreme values 
of $\theta$ remain pathological for the original canonical purification, and to a lesser
extent for the HPCP. One solution would be to break the symmetry of the McWeeny function 
by moving $\xpm$ towards $\xpp$ or $\xph$ depending on the $\theta$ value. 
Basically, this requires a higher polynomial degree for $\funcMcW$, 
\ie\ $\Trace{(\Dm^n - \Dm)^2}_{n>2}$, resulting in a higher computational 
complexity. Assuming optimal programming, we emphasize that the PMCP and HPCP
involved only two matrix multiplications per iteration. As already proved in \citen{PM_PRB_1998}, 
and highlighted by the energy convergence profiles in \fig{figure3}b, the PMCP and HPCP 
approach the (one particle) ground-state energy $\Egs=\Trace{\Hmgs\Dmgs}$ monotonically,
in other words, they are variational with respect to the Lagrange multiplier $\gamma$.
The dependence of $\pb$ on the band gap plotted in \fig{figure3}c 
confirms the early numerical experiments\cite{Niklasson_TCP_2002,Rudberg_JPCM_2011},
where $\pb$ increases linearly with respect to $\ln(1/\gap)$. The influence of $\theta$
is clearly apparent if we compare the minimum number of purification as required for the 
wider band gap ($y$-axis intercept),  where for example, with $\theta=0.5$, 
both canonical purifications reach the ideal value of about 10 purifications,
whereas for $\theta=0.05$, $\pb_\trm{HPCP}=23$ and $\pb_\trm{PMCP}=37$.

Let us consider how to improve the performance of the canonical purifications by working 
on the initial guess, regarding the hole-particle equivalence (or duality\citep{Mazziotti_PRE_2003}).
Instead of searching for $\Dm$, we may choose to purify $\Db$,
which simply requires replacing $\Dm$ with $\Db$ in the relation \reff{eq:hpcp3}.
In that case, the initial hole density matrix, satisfying $\lambda(\Db_0)\in[0,1]$, would 
be given by Eqs.~\reff{eq:dmguess} and \reff{eq:pmcpguess}, with $\beta_1 = \thetab$ and 
$\beta_2 = -\max\left\lbrace\tbeta,\bbeta\right\rbrace$. Then, intuitively, the guess for 
the particle density matrix should be improved by using this additional information.
Therefore, a more general preconditioning is proposed:
\begin{equation}
\label{eq:hpcpguess}
	\Dmhp_0 = \alpha\Dm_0 + (1-\alpha)(\Im - \Db_0)
\end{equation}
where $\alpha$ can be view as a mixing coefficient\footnote{In that case it can be
shown that: $\lambda(\Dmhp_0)\in[-\frac{1}{2},\frac{3}{2}]$}.
Results obtained with this new precontionning are plotted in \fig{figure3} 
(notated PMCP+ and HPCP+). As evident from \fig{figure3}a, the  naive value of $\alpha=0.5$ 
leads to a net improvement of the PMCP and HPCP performances over the range 
$0.3<\theta<0.7$, inside of  which the number of purifications becomes independent
of $\theta$. Outside this interval, runaway solutions were encountered due to the 
ill-conditioning of $c$, where either of the constraints in~\eq{eq:pmcpcons2} 
or~\reff{eq:pmcpcons3} is violated. The solution to this problem
is to perform a constrained search of $\alpha$ in \eq{eq:hpcpguess}, 
such that the first inequality of \eq{eq:pmcpcons2} is respected, that is:
\begin{equation}
\label{eq:hpcpguessalpha}
	\underset{
	\begin{subarray}{c}	
	0\leq\alpha\leq 1\\ 
	\delta>0
	\end{subarray}}{\operatorname{search}}\: 
	\left\lbrace  \Trace{\Dm^2_0}  =
	\begin{cases}
		\Ne-\delta\Ne,\;\text{if}\: \theta   < (1-\delta)  \\ 
    	\Ne-\delta\Nb,\;\text{if}\: \theta   > (1-\delta)  
   	\end{cases}
  \right\rbrace
\end{equation}
which leads to solve a second-order polynomial equation in $\alpha$, at the extra cost
of only one matrix multiplication. Obviously, the parameter $\delta$ has to be carefully chosen 
such that the second equality of \eq{eq:pmcpcons2} and condition \reff{eq:pmcpcons3} are also 
respected. We found $\delta \simeq 2/3$ as the optimal value\cite{MRD_unpublished}.
From \fig{figure3}, the benefits of this optimized preconditioning are clear when focussing within the 
range $]0.0,0.3]\cup[0.7,1.0[$, albeit with one or two extra purifications around the poles 
$\theta=\{0.3,0.7\}$ required to achieve the desired convergence. These benefits are even 
clearer in \fig{figure3}c, where we also show the plots of 
$\pb$ as a function of $\ln(1/\gap)$ for the test case $\theta=0.01$. At the intercept
we find $\pb_\trm{PMCP}\simeq 38$ compared to  $\pb_\trm{HPCP}\simeq 21$, 
showing the improvment bring by the hole-particle equivalence.
We have also compared our method against the most efficient of the trace updating 
methods, TRS4\cite{Niklasson_TRS_2003}, and find that for non-pathological fillings, 
the two are comparable in efficiency. For the pathological cases, where TRS4 adjusts 
the polynomial, it is more efficient, but at the expense of non-variational behaviour
in the early iterations.  

To conclude, we have shown how, by considering both electron and hole occupancies, 
the density matrix for a given system can be found efficiently while preserving $N$-representability.  
This opens the door to more robust, stable ground state minimisation algorithm, with application 
to standard and linear scaling DFT approaches.

\begin{acknowledgments}
LAT would like to acknowledge D. Hache for its unwavering support and midnight talks about 
how to move beads along a double-well potential.
\end{acknowledgments}
\clearpage
\appendix
\section{Alternative derivation of the \textit{hole-particle} canonical purification}
\label{appendixA}
We demonstrate that by symmetrizing the Palser and Manolopoulos (PM) relations
[Eqs. (16) of \citen{PM_PRB_1998}] with respect to the hole density matrix, the closed-form 
of \eq{eq:hpcp3} appears naturally. Throughout the demonstration, quantities related to 
unoccupied subspace are indicated by a bar accent. Let us start from PM equations:
\begin{subequations}
\label{eq:pm1}
\begin{align}
	\trm{for}\ & \cn\leq\frac{1}{2}: \label{eq:pm1a}\\
	&\Dm_{n+1}  = - \frac{1}{1-\cn}\Dm_n^3 
	+ \frac{1+\cn}{1-\cn}\Dm_n^2
	+ \frac{1-2\cn}{1-\cn}\Dm_n\nonumber\\
	\trm{for}\ & \cn >  \frac{1}{2}:  \label{eq:pm1b}\\  
	&\Dm_{n+1}  = - \frac{1}{\cn}\Dm_n^3 
	+ \frac{1+\cn}{\cn}\Dm_n^2\nonumber            
\end{align}
\end{subequations}
with $\cn$ given in \eq{eq:mcwgamb}. 
We may search for purification relations \textit{dual} 
to \eq{eq:pm1}, \ie\ function of $\Db$. We obtain:
\begin{subequations}
\label{eq:pm2}
\begin{align}
	\trm{for}\ & \cnb \geq \frac{1}{2}: \label{eq:pm2a} \\
	&\Db_{n+1} = - \frac{1}{1-\cnb}\Db_n^3 
                  + \frac{1+\cnb}{1-\cnb}\Db_n^2  
                  + \frac{1-2\cnb}{1-\cnb}\Db_n\nonumber\\                
	\trm{for}\ &\cnb <  \frac{1}{2}: \label{eq:pm2b} \\
	&\Db_{n+1} = -\frac{1}{\cnb}\Db_n^3 
                  + \frac{1+\cnb}{\cnb}\Db_n^2\nonumber             
\end{align}
\end{subequations}
with $\cnb = 1 - \cn$. Instead of purifying either $\Dm$ or $\Db$, 
we shall try to take advantage of the closure relation [\eq{eq:closure}] in such a way that, 
if we choose to work within the subspace of occupied states, the purification 
of $\Dm$ [\eq{eq:pm1}] is constrained to verify $\Dm = \Im - \Db$. 
By inserting this constraint in \eq{eq:pm2}, we obtain:
\begin{widetext}
\begin{subeqnarray}
\slabel{eq:pm3a}
\trm{for}\ \ \cn\leq\frac{1}{2}: & &\  \  \Dm_{n+1}  = \I -\left(-\frac{1}{\cn}(\I-\Dm_n)^3\
                  + \frac{2-\cn}{\cn}(\I-\Dm_n)^2 - \frac{1-2\cn}{\cn}(\I-\Dm_n)\right)\\  
\slabel{eq:pm3b}
\trm{for}\ \ \cn >  \frac{1}{2}: & &\  \  \Dm_{n+1} = \I-\left(-\frac{1}{1-\cn}(\I-\Dm_n)^3 
                  + \frac{2-\cn}{1-\cn}(\I-\Dm_n)^2\right)
\end{subeqnarray}
\end{widetext}
On multiplying Eqs.~\reff{eq:pm1a} and \reff{eq:pm3a} by $(1-\cn)$ and $\cn$, respectively 
[or multiplying \eq{eq:pm1b} and \reff{eq:pm3b} by $\cn$ and $(1-\cn)$], and adding, 
we obtain:
\begin{subeqnarray}
   \Dm_{n+1}  &=& \Dm_{n} + 2\left(\Dm^2_{n}\Db_n - \cn\Dm_{n}\Db_n\right)
\end{subeqnarray}
%

%

\end{document}